\documentclass[conference]{IEEEtran}
\IEEEoverridecommandlockouts

\usepackage{graphicx}
\usepackage{subfigure}
\usepackage{xcolor}
\usepackage{url}
\usepackage{nameref}
\usepackage{hyperref}
\usepackage{diagbox}
\usepackage{adjustbox}

\usepackage{array}
\usepackage{comment}
\usepackage{graphbox}
\usepackage{pbox}
\usepackage{booktabs}
\usepackage{multirow}
\usepackage{slashbox}
\usepackage{diagbox}
\usepackage{etoolbox}
\usepackage[ruled]{algorithm2e} 
\usepackage{enumitem}
\setlist{leftmargin=3.5mm}

\def\BibTeX{{\rm B\kern-.05em{\sc i\kern-.025em b}\kern-.08em
    T\kern-.1667em\lower.7ex\hbox{E}\kern-.125emX}}
\begin{document}

\title{Micromobility in Smart Cities: A Closer Look at Shared Dockless E-Scooters via Big Social Data}

\author{\IEEEauthorblockN{Yunhe Feng}
\IEEEauthorblockA{\textit{Information School} \\
\textit{University of Washington}\\
Seattle, USA \\
yunhe@uw.edu}
\and
\IEEEauthorblockN{Dong Zhong}
\IEEEauthorblockA{\textit{Department of EECS} \\
\textit{University of Tennessee}\\
Knoxville, USA \\
dzhong@vols.utk.edu}
\and
\IEEEauthorblockN{Peng Sun}
\IEEEauthorblockA{\textit{Shenzhen Institute of Artificial Intelligence and Robotics for Society} \\
\textit{The Chinese University of Hong Kong, Shenzhen}\\
Guangdong, China \\
sunpengzju@gmail.com}
\and
\IEEEauthorblockN{Weijian Zheng}
\IEEEauthorblockA{\textit{Department of Computer Science} \\
\textit{Purdue University}\\
West Lafayette, USA \\
zheng273@purdue.edu}
\and
\IEEEauthorblockN{Qinglei Cao}
\IEEEauthorblockA{\textit{Department of EECS} \\
\textit{University of Tennessee}\\
Knoxville, USA \\
qcao3@vols.utk.edu}
\and
\IEEEauthorblockN{Xi Luo}
\IEEEauthorblockA{\textit{SLAC National Accelerator Laboratory} \\ \\
San Mateo, USA \\
xluo@slac.stanford.edu}
\and
\IEEEauthorblockN{Zheng Lu}
\IEEEauthorblockA{\textit{Department of EECS} \\
\textit{University of Tennessee}\\
Knoxville, USA \\
zlu12@vols.utk.edu}
}

\maketitle

\begin{abstract}
The micromobility is shaping first- and last-mile travels in urban areas.
Recently, shared dockless electric scooters (e-scooters) have emerged as a daily alternative to driving for short-distance commuters in large cities due to the affordability, easy accessibility via an app, and zero emissions.
Meanwhile, e-scooters come with challenges in city management, such as traffic rules, public safety, parking regulations, and liability issues.
In this paper, we collected and investigated 5.8 million scooter-tagged tweets and 144,197 images, generated by 2.7 million users from October 2018 to March 2020, to take a closer look at shared e-scooters via crowdsourcing data analytics. 
We profiled e-scooter usages from spatial-temporal perspectives, explored different business roles (i.e., riders, gig workers, and ridesharing companies), examined operation patterns (e.g., injury types, and parking behaviors), and conducted sentiment analysis.
To our best knowledge, this paper is the first large-scale systematic study on shared e-scooters using big social data.
\end{abstract}

\begin{IEEEkeywords}
ridesharing, big data, crowdsourcing, smart city, Twitter, shared e-scooter, micromobility, micro-mobility, scooter, transportation, social networks
\end{IEEEkeywords}

\section{Introduction}

Micromobility is an emerging term usually referring to the usage of docked and dockless lightweight devices (e.g., bikes) for short- and medium-length trips.
As a new mode of micromobility, shared dockless electric scooters (e-scooters) are gaining popularity in recent years.
A recent survey conducted in February 2019 showed 11\% of Paris residents reported using e-scooters either frequently or from time to time~\cite{odoxa-report}.
Aiming at closing first- and last-mile transit gaps for residents, many ridesharing companies, such as \textit{Lime}, \textit{Bird}, and \textit{Lyft}, deployed thousands of e-scooters in more than 60 cities across the United States.
According to the National Association of City Transportation Officials (NACTO)~\cite{NACTO-report}, people took 38.5 million trips on shared e-scooters in 2018.

\begin{figure}[ht]
\subfigure[E-scooter parked outside a plaza
  \label{fig:scooter_demo}]{\includegraphics[width=0.3\linewidth]{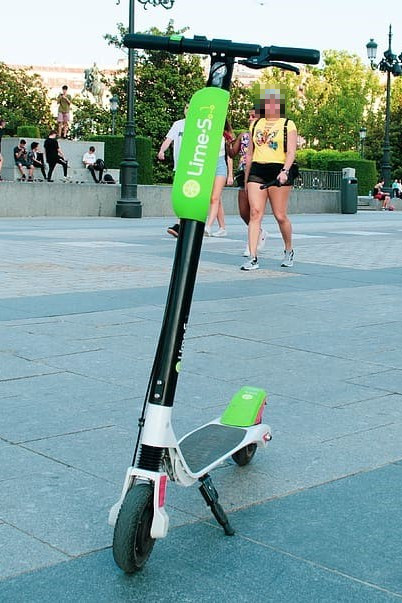}}
\subfigure[E-scooter apps
    \label{fig:scooter_apps}]{\includegraphics[width=0.3\linewidth]{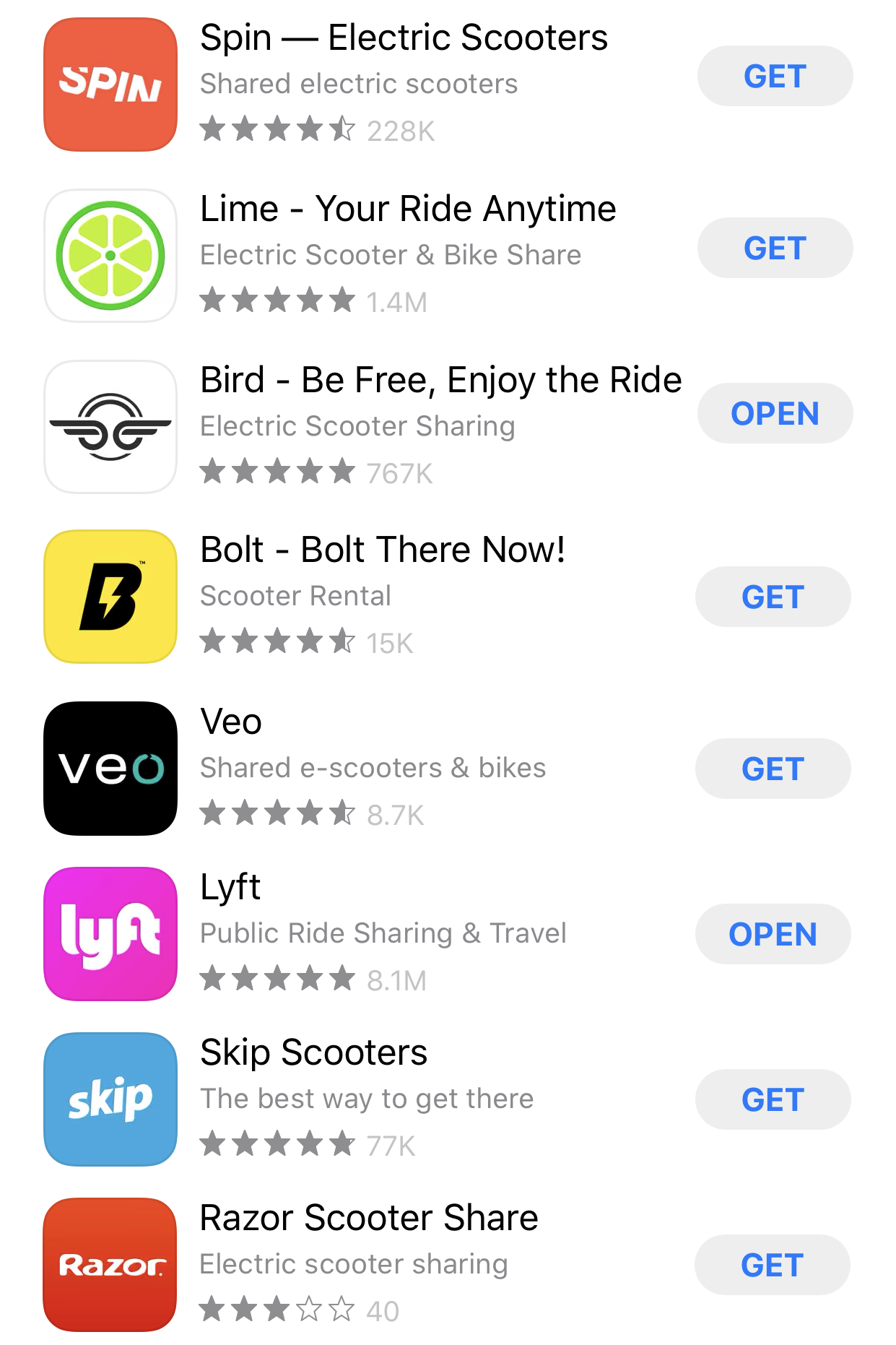}}
\subfigure[User interface of e-scooter apps
    \label{fig:scooter_UI}]{\includegraphics[width=0.3\linewidth]{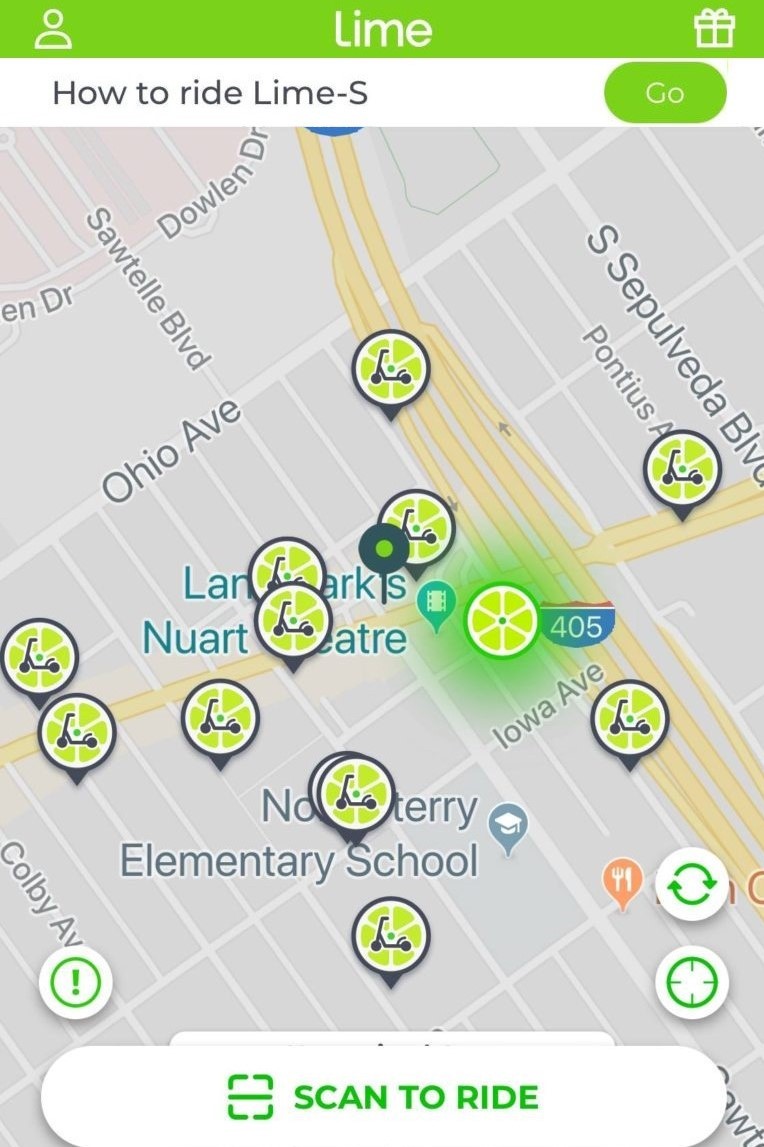}}
  \caption{Shared dockless e-scooters and smartphone apps}
  \label{fig:scooter_demo_and_UI}
\end{figure}

Smartphones are one of the key enablers of e-scooter sharing service.
To ride a shared e-scooter, users must download e-scooter apps (see Figure~\ref{fig:scooter_apps}), sign up, input payment information, and scan a QR code to unlock the e-scooter.
Figure~\ref{fig:scooter_demo} shows an e-scooter parked outside one plaza and Figure~\ref{fig:scooter_UI} illustrates a user interface of e-scooter apps.
On the app, riders can check ready-to-go e-scooters parked nearby.
After finishing the trip, riders make a payment on the app, and the e-scooter is locked automatically.
However, most existing e-scooter studies ignored the feature that e-scooters must be operated through smartphones, leading to a missing research perspective from the riders' comments shared via smartphones, especially via social media apps.

We think social media is a good data source to investigate the e-scooter usages because of its diversity, scalability, and transparency~\cite{feng2019chasing}.
First, the multimodal data available on social media enables detailed profiles of e-scooter sharing services in various aspects.
For example, free-form text can infer topics that people care about.
The shared images can be used to analyze gender gaps and self-reported injuries of riders. 
The posted timestamps and GPS information make the temporal-spatial analysis possible.
Even the embedded emojis contribute to the sentiment analysis.
Second, it is flexible and effortless to expand to a large scale regarding when the survey is conducted, how long it lasts, which e-scooter brands, cities, and even countries are considered.
On the contrary, interviews, questionnaires and observations based surveys, which many existing works rely on, lack such scalability.
Third, users expect the data they posted on some social media platforms (e.g., Twitter) to be publicly available, addressing potential concerns of such non-reproducibility caused by using unreachable first-party data. 

In this paper, we chose to use Twitter as our lens to examine shared e-scooters comprehensively through big social data analytics.
Specifically, we monitored and tracked 5.8 million English tweets mentioning the word ``scooter'' or the scooter emoji~\includegraphics[width=0.023\linewidth]{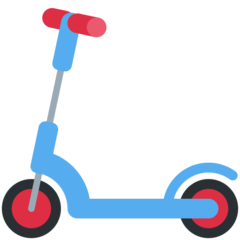} via the Twitter Streaming APIs in a real-time manner from October 6, 2018 to March 14, 2020.
After cleaning data, we presented an overview of temporal (both monthly and hourly) and geospatial tweet distributions.
Then, we extracted the involved popular topics using both social media exclusive tools (i.e., \#hashtags) and general topic models.
The discovered topics could be summarized into four categories, i.e., e-scooter deployments, stakeholders, operations, and emotions.
For topics in each category, we leveraged heterogeneous Twitter data, including text, @mentions, GPS data, general photos, screenshots of e-scooter apps, emojis, and emoticons, to reveal useful patterns.

As the first step to conduct a systematic, large-scale study on shared e-scooter using big social data, contributions and findings of this paper can be summarized as follows:
\begin{itemize}
    \item Trends of scooter usages indicated by the number of tweets varied from country to country: a decreasing pattern for the United States, New Zealand, and Australia, stable for the United Kingdom, and increasing for Canada and India.
    \item We inferred twelve topics people discussed extensively on Twitter, such as shared e-scooter regulations in cities, gig jobs, parking issues, and scooter-related injuries.
    \item We profiled geospatial distributions of e-scooter tweets at the city level across the United States, and summarized the commonalities of local regulations on shared e-scooters.
    \item Using both tweet text and brand logos recognized automatically from images, we illustrated e-scooter market shares.
    \item We confirmed a gender gap in shared e-scooter riders with 34.86\% identified as female and 65.14\% as male.
    \item We estimated the median trip payment and duration by e-scooter app screenshots, and classified scooter-related injuries and parking places.
    \item We also conducted a comprehensive social sentiment analysis via polarized words, facial emojis, and emoticons to measure the general public's emotions and feelings on e-scooter sharing services.
\end{itemize}

\section{Related Work}

There is a large body of work investigating the advantages, disadvantages, and problems of shared e-scooters in urban transportation.
Severengiz et al.~\cite{severengiz2020assessing} quantified the environmental impact of shared e-scooters in the city of Bochum, Germany, and demonstrated e-scooters could increase the environmental benefits.
However, another study~\cite{hollingsworth2019scooters} conducted a Monte Carlo analysis and showed that e-scooters might potentially increase life cycle emissions relative to the transportation modes that they substituted.
H{\'e}lie et al.~\cite{moreau2020dockless} reported dockless e-scooters needed a lifespan of at least 9.5 months to be a green micromobility solution.

Multiple studies have explored the challenges caused by the increased usage of shared e-scooters in urban areas. 
Bresler et al.~\cite{bresler2019craniofacial} examined the patterns of the motorized scooter related injuries for riders, and pleaded requirements to develop appropriate public policies such as using helmets to mitigate injuries.
Sikka et al.~\cite{sikka2019sharing} studied the safety risks and incidence of injuries for pedestrians who shared the sidewalk with e-scooters.
In~\cite{vinayaga2020security}, researchers revealed and summarized the potential privacy and security challenges and concerns related to e-scooters, which was helpful to both riders and service providers.

To ensure traffic safety and improve urban planning, a few recent studies have sought to enforce regulations and build public infrastructures for shared e-scooters.
G{\"o}ssling~\cite{gossling2020integrating} analyzed local media reports and concluded that urban planners needed to introduce policies regarding maximum speeds, mandatory use of bicycle lanes, and the max number of licensed operators.
Kondor et al.~\cite{kondor2019estimating} demonstrated that actual benefits brought by e-scooters highly depended on the availability of dedicated infrastructure.
McKenzie~\cite{mckenzie2019spatiotemporal} compared the spatial-temporal trip patterns between dockless e-scooters and docked bike-sharing services to offer suggestions on public policies and transportation infrastructures for e-scooters.

When analyzing the shared e-scooter usage, most of above works only focused on a particular aspect, such as environmental impacts~\cite{severengiz2020assessing,hollingsworth2019scooters,moreau2020dockless}, injuries~\cite{bresler2019craniofacial,sikka2019sharing}, security concerns~\cite{vinayaga2020security}, and infrastructure organization~\cite{gossling2020integrating,kondor2019estimating,mckenzie2019spatiotemporal}.
In this paper, we harvested millions of tweets covering 18 months to provide a comprehensive understanding of shared e-scooters.
Diverse techniques like natural language processing (NLP), optical character recognition (OCR), logo detection and recognition, and sentiment analysis were applied on heterogeneous Twitter data (e.g., text, GPS data, images, and emojis) to produce insights and patterns from multiple perspectives.

\section{Dataset}
In this section, we first described the data collection and cleanup.
Then we presented an overview of spatial-temporal distributions of the preprocessed data.

\subsection{Data Collection}

We utilized Twitter's Streaming APIs to crawl real-time tweets containing either the word \textit{scooter} or the scooter emoji~\includegraphics[width=0.023\linewidth]{figures/emoji_image/scooter_1f6f4.png}.
We collected more than 5.8 million tweets generated by 2.7 million unique users from October 6, 2018 to March 14, 2020.
We also extracted 178,048 different image URLs inserted in the collected tweets.
Among them, 144,197 images were retrieved successfully and the rest 33,851 images were expired.

\subsection{Data Cleaning}

One of the challenges when dealing with messy text like tweets is to remove noise from data.
In our study, we performed three types of noise removal to enhance data analysis step by step.
First, we detected and deleted tweets generated by Twitter bots.
Inspired by the bot detection approach proposed in \cite{ljubevsic2016global}, we conceived the two types of Twitter users as bots: (1) those who posted more than 525 scooter-tagged tweets, i.e., more than one such tweets per day during our data collection period; (2) those who posted over 100 scooter-tagged tweets in total and the top three frequent posting intervals covered at least their 90\% tweets.
For the two types of bots, we removed 104,739 tweets created by 90 bots and 8,318 tweets from 18 bots respectively.

When analyzing word occurrences, we observed that \textit{Braun}, \textit{Taylor}, \textit{Scott}, \textit{Justin}, and \textit{Swift} were among the top 20 words with the highest frequency of appearing together with \textit{scooter} in the same tweet.
After careful reviews, we found the \textit{scooter} in such tweets might not refer to the real scooter studied in this paper.
Instead, it implied Scott Samuel ``Scooter'' Braun, an American entrepreneur who triggered many hot topics with other celebrities on social media.
Therefore, we removed scooter-tagged tweets that contained the words of \textit{Taylor}, \textit{Swift}, \textit{Justin}, \textit{Bieber}, \textit{Scott}, \textit{Samuel}, \textit{Braun}, \textit{Ariana}, \textit{Grande}, and \textit{Borchetta}.
Thus, 1,541,815 related tweets were deleted.

To further reduce false-positive errors, we designed a set of keywords to distinguish shared e-scooters from other types of scooters, such as kick scooters and motor scooters.
Specifically, we picked out tweets containing at least one word of \textit{Share} and the shared e-scooter brands including \textit{Bird}, \textit{Lime}, \textit{Spin}, \textit{Bolt}, \textit{gruv}, \textit{Lyft}, \textit{Sherpa}, \textit{VeoRide}, \textit{Taxify}, \textit{Jump}, \textit{RazorUSA}, \textit{Scoot Networks}, and \textit{Skip}.
Note that our approach is flexible enough to add new shared e-scooter startups for future investigations.
Finally, we put together 416,291 tweets including original tweets, replies, retweets, quoted tweets, and 258,495 unique tweets after deleting retweets.
Besides, we obtained 17,695 images posted along with these tweets.

\subsection{Temporal Distribution}

\begin{figure*}[ht]
  \centering
  \subfigure[Monthly distribution by country 
  \label{fig:distribution_in_month_by_country}]{\includegraphics[width=0.32\linewidth]{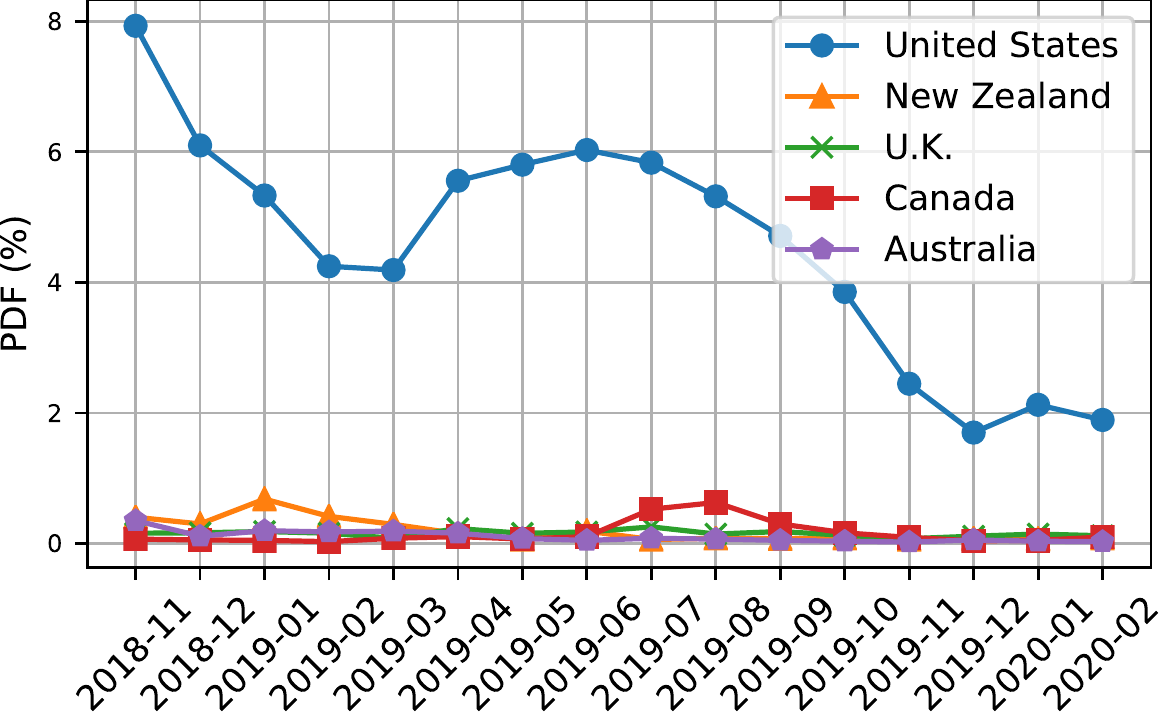}}
  \subfigure[Monthly dist. w/o United States
  \label{fig:distribution_in_month_by_country_non_US}]{\includegraphics[width=0.32\linewidth]{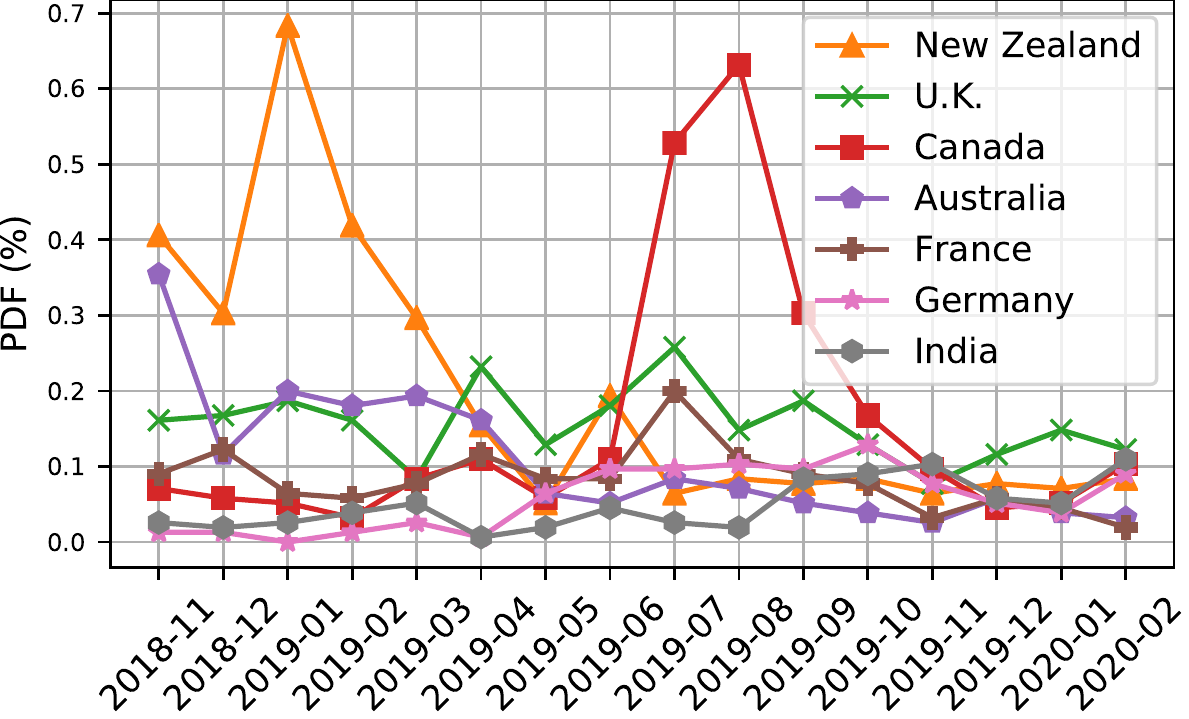}}
  \subfigure[Hourly dist. in United States
  \label{fig:distribution_by_hour}]{\includegraphics[width=0.3\linewidth]{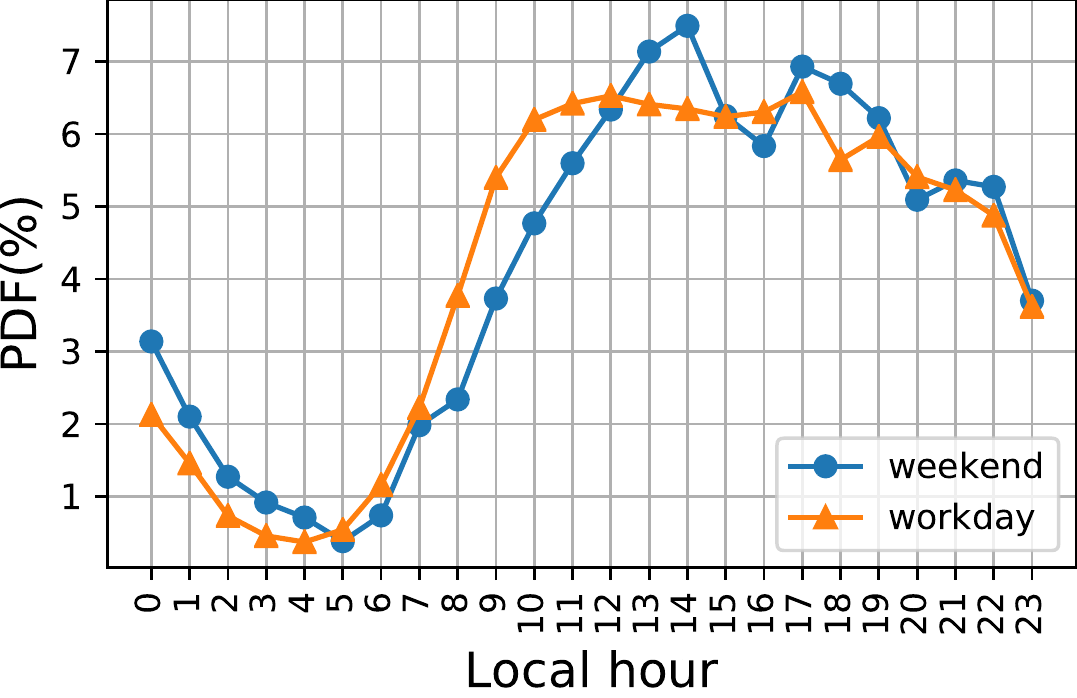}}
  \caption{Time distribution from Oct. 6, 2018 to Mar. 14, 2020. The data of Oct. 2018 and Mar. 2020 are not plotted in (a) and (b) due to the data incompleteness. 
  \label{fig:distribution_time}}
\end{figure*}

We profiled the temporal distributions of e-scooter related tweets using two time granularities, i.e., by month and by hour.
Figure~\ref{fig:distribution_in_month_by_country} and~\ref{fig:distribution_in_month_by_country_non_US} illustrate the monthly percentage of posted tweets by different countries.
Followed by New Zealand, United Kingdom, Canada, and Australia, the United States accounts for more than 82\% of all collected tweets.
As shown in Figure~\ref{fig:distribution_in_month_by_country}, months in summer contributed the highest monthly data volumes in the United States in 2019.
There is a significant drop when comparing data volumes in Nov. and Dec. 2019 with that in Nov. and Dec. 2018.
We think such a drop may be caused by two reasons.
First, e-scooter users are more likely to discuss their riding experience online when trying e-scooters for the first time.
Second, strict e-scooter regulations and policies, such as limiting the number of companies authorized to operate scooters in each city, were imposed in many U.S. cities in 2019.

Besides the United States, we investigated the monthly data distributions of other seven countries in Figure~\ref{fig:distribution_in_month_by_country_non_US}.
The two peaks in New Zealand and Canada occur in local summer months, which may indicate that e-scooters are used more frequently during summer.
Together with New Zealand, Australia shows a decreasing trend regarding the monthly data volume.
On the contrary, Germany and India demonstrate an increasing trend.
The amount of e-scooter tweets posted per month from United Kingdom and France are relatively stable.

We also explored the hourly tweet distributions regarding both workdays and weekends in the United States, as shown in Figure~\ref{fig:distribution_by_hour}.
As we expected, the tweet amount on each day of the week is lower between 0:00 am and 7:00 am than the daytime.
The most active time during weekdays is between 10:00 am to 5:00 pm (see the orange line).
However, the peak time on weekends is between 12:00 pm and 7:00 pm (see the blue line).
One possible reason is that many riders tend to start their outdoor activities later on Saturday and Sunday. 

\subsection{Geospatial Distribution}
\begin{figure}[ht]
\subfigure[Percentage by state 
  \label{fig:distribution_in_space_US_map}]{\includegraphics[width=0.49\linewidth]{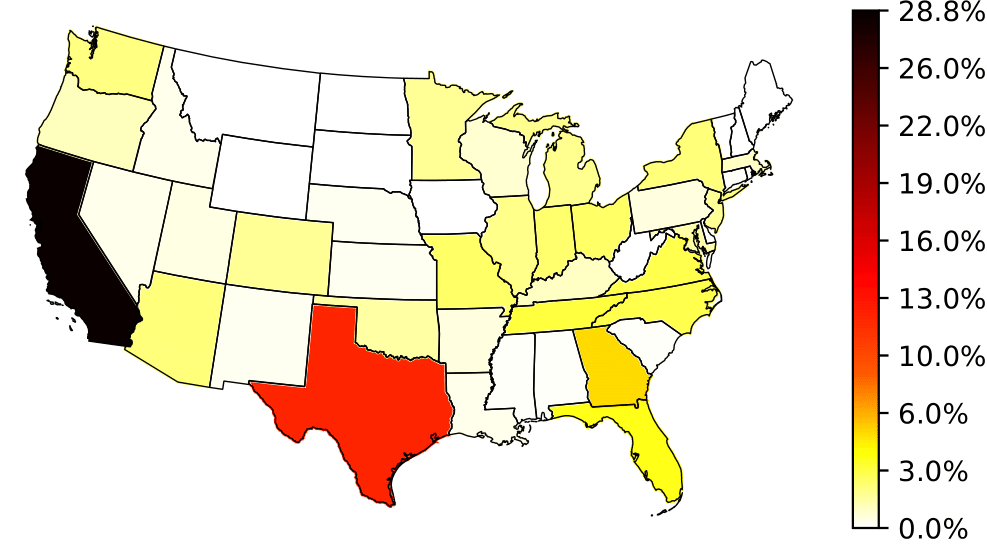}}
\subfigure[Normalized by state population
  \label{fig:distribution_in_space_US_map_normalized}]{\includegraphics[width=0.49\linewidth]{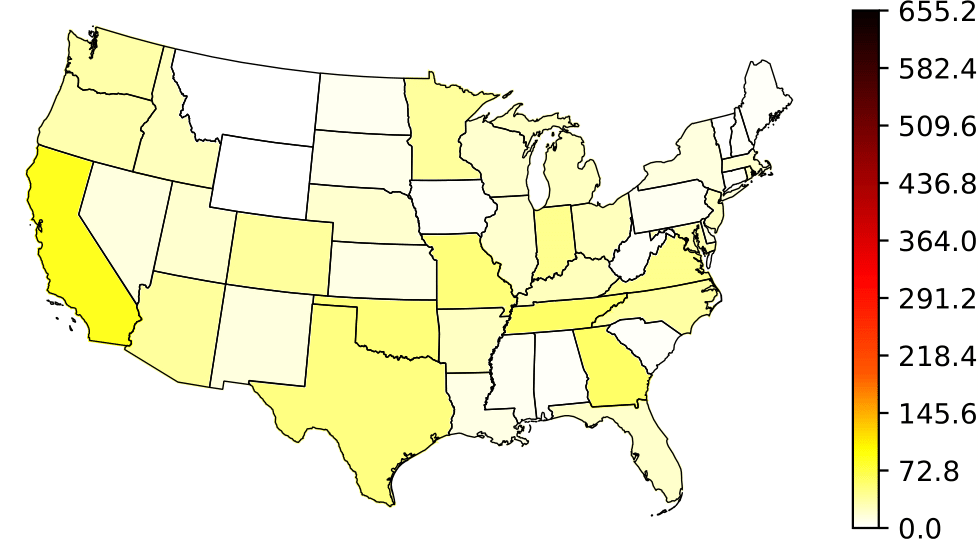}}
  \caption{Geospatial distribution across United States}
  \label{fig:distribution_in_space_US_map_all}
\end{figure}

We selected the United States as an example to study the geospatial distribution of e-scooter related tweets at the state level.
The percentage of tweets posted from each state in the United States is demonstrated in Figure~\ref{fig:distribution_in_space_US_map}, where California (28.8\%) and Texas (11.7\%) account for more than 40\% of all tweets.
We also noticed six (CA, TX, GA, FL, NC, OH) out of the top ten states with the highest percentages were among the ten most populous states.
After normalizing by state population, we obtained a relatively smooth distribution, as shown in Figure~\ref{fig:distribution_in_space_US_map_normalized}.
It is interesting to note that Washington D.C., one of the least populated states, generated the highest number of posted tweets per million residents.
McKenzie~\cite{mckenzie2019spatiotemporal} reported that scooter-share trips in Washington D.C. supported leisure, recreation, or tourism activities more than commuting, which may explain our findings.

\section{Topic Discovery}
In this section, we explored and summarized underlying topics about e-scooter sharing services on social media.
We mainly applied two methodologies, i.e., \#Hashtags and Latent Dirichlet Allocation (LDA) topic modeling to discover and summarize what Twitter users discussed online.

\subsection{Hashtags Analysis}
\begin{figure}[ht]
  \includegraphics[width=\linewidth]{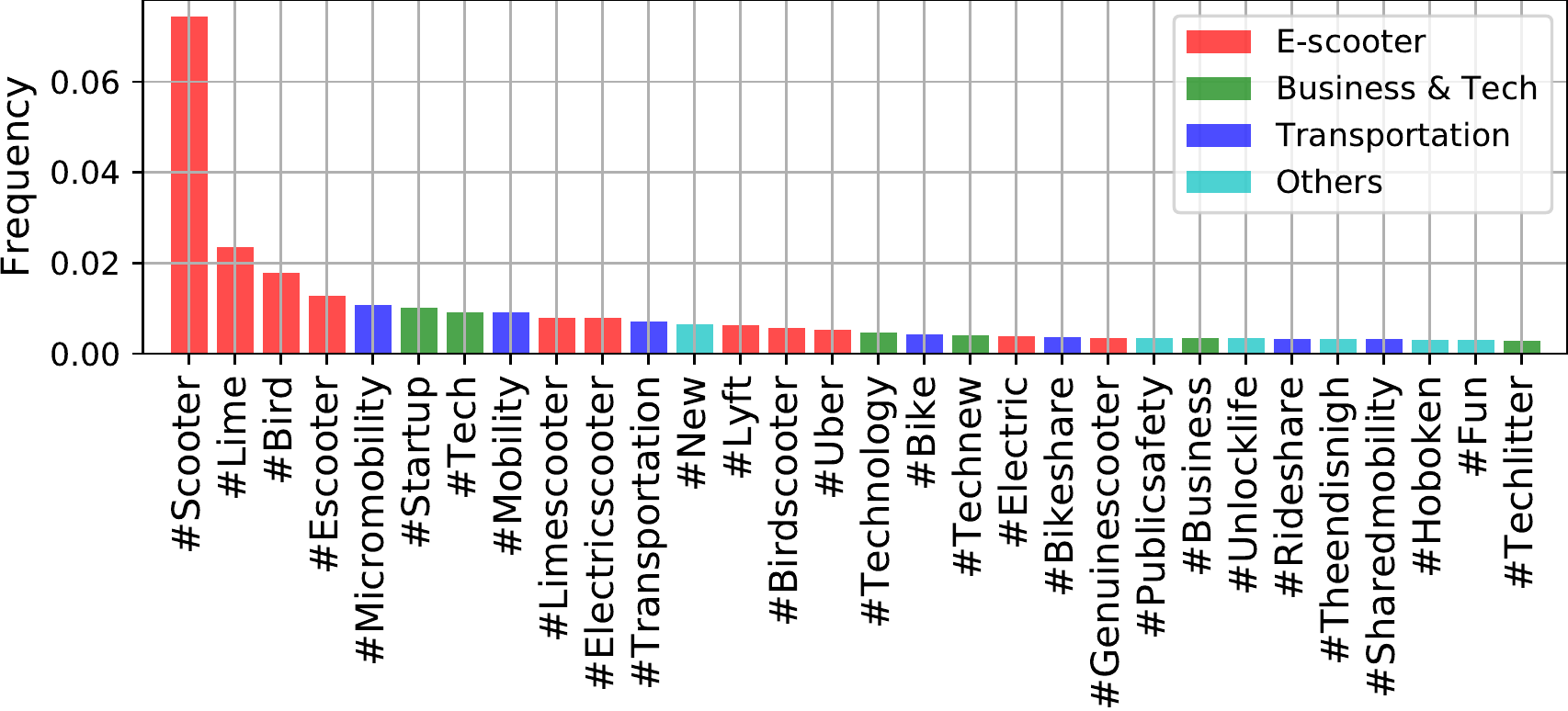}
  \caption{Hashtag-based topics
  \label{fig:hashtags}}
\end{figure}

The \#hashtags play a crucial role in indexing keywords and discussing specific topics.
According to Twitter, hashtagged words that become very popular are often trending topics~\cite{tweet-hashtag}.
We observed and extracted more than 25,800 unique hashtags in our dataset.
The 25 most popular hashtags are illustrated in Figure~\ref{fig:hashtags}.
Unsurprisingly, \textit{\#scooter} was the most frequently mentioned hashtag because it is the topic that this paper focuses on. 
We grouped all those hashtags into four topics, namely e-scooter, business, transportation, and others.

The topic of business consists of shared e-scooter brands including \textit{\#Lime, \#Bird, \#Lyft,} and \textit{\#Uber}, which implied that such startups were playing a key role in e-scooter ridesharing business. 
In the topic of transportation, hashtags like \textit{\#Micromobility}, \textit{\#Mobility}, and \textit{\#Rideshare} demonstrated shared e-scooters was indeed an important alternative mode in micromobility.
People also used hashtags of \textit{\#Tech, \#New, \#Technew, \#Unlocklife} to describe e-scooters, showing this new technology was changing the life.
Meanwhile, people expressed their concerns on the \textit{\#PublicSafety} threatened by shared e-scooters. 

\subsection{Topic Model}

The Latent Dirichlet Allocation (LDA) is one of the most widely used topic models in text mining to gain deep and meaningful insights from unstructured data.
We treated each tweet as an individual document to build a corpus to train the LDA.
On each document, we first filtered out commonly used stop words, then tokenized, lemmatized, and stemmed the rest words.
On the entire corpus, we applied the Term Frequency-Inverse Document Frequency (TF-IDF) to drop irrelevant words and give high weights to important ones.

When tuning LDA, it is challenging to determine the best number of topics.
In our study, we utilized $C_v$ metric, which was reported as the best coherence measure by combining normalized pointwise mutual information (NPMI) and the cosine similarity~\cite{roder2015exploring}, to evaluate the difference between topics and the similarity inside each topic.
We trained LDA models with the number of topics ranging from 1 to 20 for 500 iterations respectively.
We found the highest coherence score was achieved when setting the number of topics as $12$.

As the LDA model produced a list of words representing each topic, we manually parsed the word lists and assigned topic names accordingly.
Table~\ref{tab:topic_model} shows the 12 topics we concluded based on the LDA results.
These topics were further clustered into four categories, namely Deployment, Stakeholder, Operation, and Emotion, based on their meanings and domains.
Specifically, we grouped Transportation, City, and Regulation into the category of Deployment because these topics are related to e-scooter deployment.
Three distinct roles in business, i.e., e-scooter riders, gig workers (such as chargers), and e-scooter operating companies, formed the category of Stakeholder.
We also identified four typical aspects that people were concerned about during scooter operations, including scooter products, transactions, parking, and injuries associated with e-scooters.
At last, both positive and negative emotions involved in shared e-scooters were categorized as Emotion.
In the following sections, we will present detailed insights for each category.

\begin{table*}[]
\centering
\setlength{\tabcolsep}{3pt}
\begin{adjustbox}{
  addcode={\begin{minipage}{\width}
    \caption{The extracted topics using the LDA topic model}
    \label{tab:topic_model}}
    {\end{minipage}},
    rotate=0}
\centering
\begin{tabular}{@{}l|lll|ll|lll|llll@{}}
\toprule
              & \multicolumn{3}{c|}{\textbf{Stakeholder}}                      & \multicolumn{2}{c|}{\textbf{Emotion}}  & \multicolumn{3}{c|}{\textbf{Deployment}}                          & \multicolumn{4}{c}{\textbf{Operation}}                                       \\ \midrule
\textbf{Rank} & \textbf{Rider}   & \textbf{Gig Worker}  & \textbf{Company} & \textbf{Positive} & \textbf{Negative} & \textbf{Transport.} & \textbf{City}    & \textbf{Regulation} & \textbf{Parking} & \textbf{Transaction}  & \textbf{Injury}   & \textbf{Product}  \\ \midrule
1             & kid              & gig              & electr           & ride              & ride              & bike                    & santa            & citi                & sidewalk         & app               & injuri            & balanc            \\
2             & kick             & worker           & startup          & around            & one               & share                   & monica           & program             & park             & ride              & fire              & self              \\
3             & adjust           & built            & compani          & fun               & got               & transport               & st               & compani             & bike             & charg             & rider             & electr            \\
4             & wheel            & contractor       & san              & day               & saw               & car                     & loui             & pilot               & peopl            & use               & accid             & bo                \\
5             & height           & gen              & market           & love              & shit              & citi                    & montreal         & electr              & road             & free              & recal             & skateboard        \\
6             & amp              & economi          & via              & downtown          & time              & transit                 & paul             & new                 & lane             & unlock            & caus              & smart             \\
7             & child            & lemon            & new              & san               & fuck              & trip                    & toronto          & council             & use              & code              & man               & board             \\
8             & light            & kmh              & launch           & time              & guy               & use                     & joe              & bike                & block            & tri               & injur             & inch              \\
9             & adult            & stabl            & ford             & rode              & hit               & mobil                   & canal            & share               & ride             & one               & via               & fold              \\
10            & boy              & libbi            & tech             & one               & someon            & electr                  & german           & come                & pedestrian       & minut             & report            & bluetooth         \\
11            & new              & bigger           & share            & today             & rode              & amp                     & mike             & launch              & street           & work              & electr            & led               \\
12            & fold             & revel            & news             & great             & tri               & option                  & visual           & oper                & helmet           & bike              & auckland          & hack              \\
13            & led              & surviv           & francisco        & adventur          & today             & make                    & endtoend         & amp                 & rider            & time              & say               & 65                \\
14            & pro              & abus             & busi             & first             & go                & public                  & canada           & street              & car              & pleas             & crash             & deal              \\
15            & flash            & brooklyn         & million          & back              & peopl             & ride                    & venic            & permit              & need             & download          & compani           & speaker           \\
16            & young            & flamingo         & citi             & amp               & walk              & compani                 & dave             & regul               & amp              & map               & brake             & two               \\
17            & toy              & wellington       & technolog        & way               & almost            & new                     & sunset           & lo                  & citi             & lock              & lawsuit           & portabl           \\
18            & old              & independ         & buy              & see               & thing             & peopl                   & presid           & angel               & one              & pay               & break             & hover             \\
19            & seat             & employe          & bike             & photo             & around            & need                    & impair           & today               & way              & find              & batteri           & max               \\
20            & girl             & heavier          & oper             & thing             & night             & bu                      & print            & ban                 & make             & help              & polic             & skate             \\
21            & year             & wider            & scoot            & weekend           & see               & think                   & krau             & offici              & traffic          & money             & pull              & valley            \\
22            & toddler          & bolton           & rais             & take              & as                & vehicl                  & jason            & campu               & see              & hey               & death             & silicon           \\
23            & brake            & warsaw           & servic           & new               & first             & mode                    & spray            & safeti              & mani             & account           & safeti            & roller            \\
24            & age              & superior         & mobil            & got               & man               & great                   & pier             & approv              & think            & day               & news              & certifi           \\
25            & sport            & poland           & acquir           & thank             & jump              & mile                    & georg            & bring               & user             & locat             & glitch            & motor             \\
26            & deck             & sturdier         & fleet            & much              & lol               & replac                  & ghost            & say                 & danger           & credit            & street            & threat            \\
27            & electr           & specialist       & rental           & electr            & realli            & work                    & bankrupt         & ride                & space            & need              & dc                & mini              \\
28            & big              & neat             & entrepreneur     & enjoy             & back              & way                     & expo             & allow               & speed            & thank             & issu              & spinner           \\
29            & skate            & brad             & fund             & bird              & car               & see                     & mongoos          & join                & pleas            & servic            & woman             & hutt              \\
30            & black            & ditch            & ceo              & park              & last              & data                    & panic            & transport           & place            & take              & concern           & light             \\ \bottomrule
\end{tabular}
\end{adjustbox}
\end{table*}

\section{Scooter Deployment}
Shared e-scooters are changing public transportation, especially first- and last-mile travels in cities.
In this section, we studied the geospatial distributions of scooter-mentioned tweets at the city level, and explored the policies and regulations on shared e-scooters enforced by local authorities. 

\subsection{Deployment in Cities}
\begin{figure*}[ht]
\subfigure[Exact $(lat, lon)$ coordinates
  \label{fig:precise_geo_tagged_tweets}]{\includegraphics[width=0.49\linewidth]{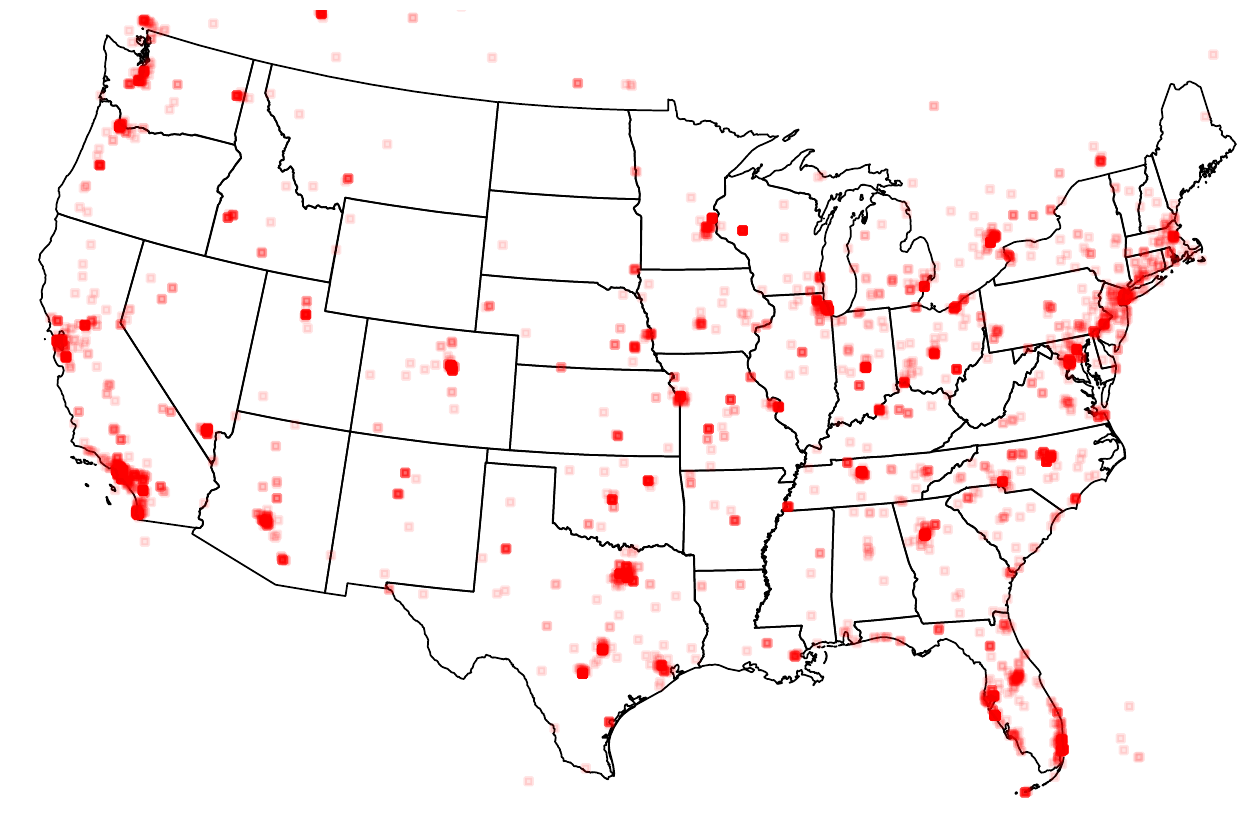}}
\subfigure[Geospatial distribution by county
  \label{fig:dict_parsed_US_coordinate}]{\includegraphics[width=0.49\linewidth]{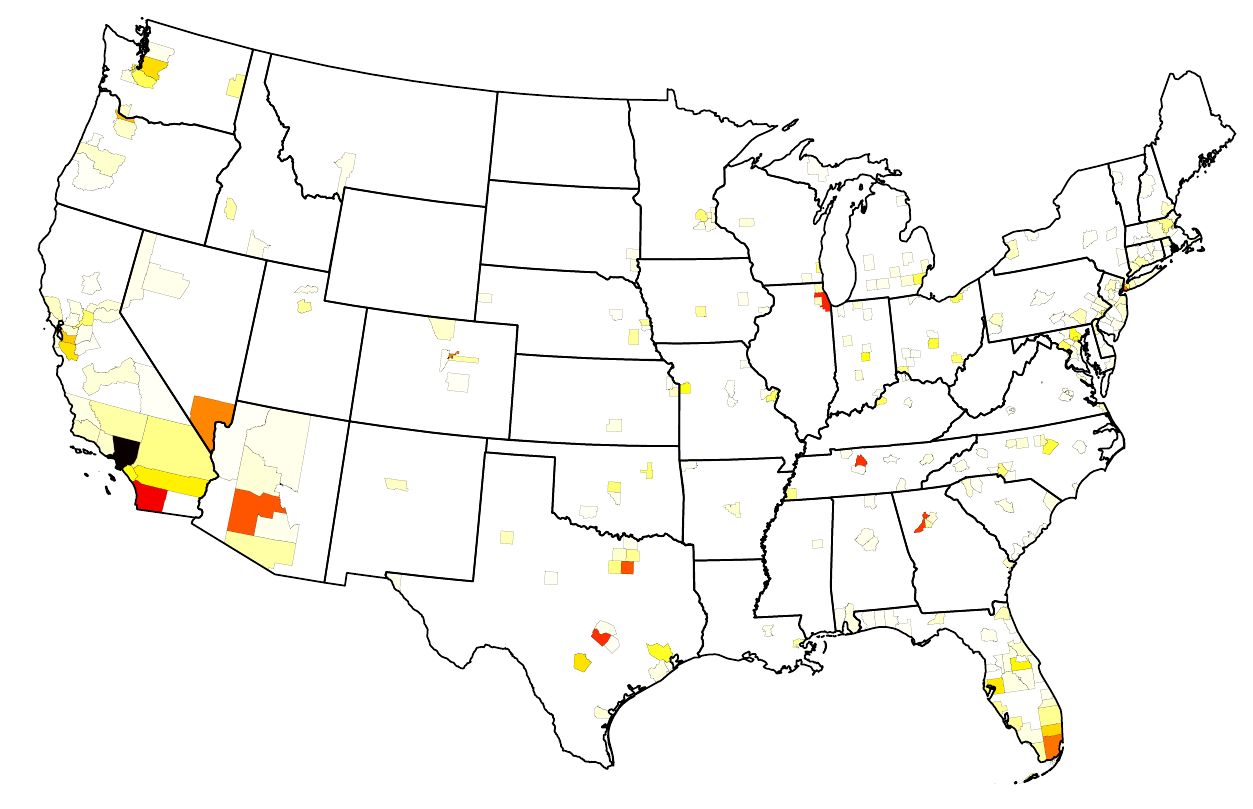}}
  \caption{Distribution of tweets with exact GPS coordinates at the city level}
  \label{fig:city_distribution}
\end{figure*}

We leveraged tweets geo-tagged with precise GPS $(lat, lon)$ coordinates to explore the scooter deployment in cities.
Within the United States, we collected 3359 exact GPS coordinates located in 579 cities.
The exact GPS coordinates are demonstrated in Figure~\ref{fig:precise_geo_tagged_tweets}, where a deeper color indicates a higher GPS data density.
Figure~\ref{fig:dict_parsed_US_coordinate} shows the GPS data distribution aggregated by county.
From the two figures, we can see that most extensive scooter deployments occurred in large cities at East Coast and West Coast, and other metropolises of the United States.
Three cities in California contributed more than 15.6\% of all GPS-tagged tweets -- Los Angeles with a proportion of 7.9\%, San Francisco (4.2\%), and San Diego (3.5\%).
Washington D.C.(2.1\%), New York (2.1\%), and Miami (1.5\%) ranked as the top three in the east coast cities.
In addition, scooters were also very popular in metropolises including Chicago (2.6\%), Austin (2.4\%), Nashville (2.3\%), Atlanta (2.2\%), Dallas (1.7\%), and Denver (1.5\%).

\subsection{Policies and Regulations}

For the topic of regulation in Table~\ref{tab:topic_model}, we surveyed policies and regulations on shared e-scooters in ten cities where the most GPS data were collected.
Although the rules in the cities are slightly different, most of them are very similar and contain keywords in our extracted regulation topic. 
We summarize those similarities into the following four categories.

\begin{itemize}
\item \textbf{Self-protection requirements}: When operating e-scooters, users are usually required to wear protective equipment such as helmets. At night, headlights and reflective-stickers are usually required.
\item \textbf{Riding behaviors}: (1) Riders cannot use any electronic devices, including the phone, while riding e-scooters. (2) There cannot be more than one rider on an e-scooter unless it is specifically designed to carry more than one person. 
\item \textbf{Traffic restrictions}: Ride e-scooters in bike lanes or sidewalks with reasonable maximum speeds in a range from 15 mph to 30 mph. 
\item \textbf{Parking rules}: Scooters cannot park in parking spaces designed for cars and in such a manner that blocks pedestrian, crosswalks, doorways, driveways or vehicle traffic.
\end{itemize}

\section{stakeholders in the Scooter Business}
In this section, we profiled three typical stakeholders in the scooter business, namely riders, gig workers, and companies.

\subsection{Riders}
Similar to other businesses, customers play a key role in the popularity of e-scooter ridesharing services.
We investigated e-scooter riders from three aspects, i.e., genders, ages, and whether wearing a helmet.
We randomly selected 10\% (1770) images from our collected image dataset and recognized 94 images containing riders.
Then, we manually conducted three classifications for the above three profiling aspects.
We found a great gender gap in shared e-scooters with 34.86\% identified as female and 65.14\% as male.
Our findings are consistent with a recent report by Portland State University~\cite{scooter-gender-gap}, which reported 34\% identified as a woman, 64\% as a man, and 2\% as transgender or non-binary.
In regards to ages, we labeled riders as either adults or kids.
It is not surprising that only 4.17\% riders were recognized as kids.
We also observed 83.51\% users did not wear a helmet when riding, which might be one of the most frequent risky behaviors.

\subsection{Gig Workers}
Before 2019, many e-scooter sharing startups pay independent contractors, i.e., gig workers, to help with the operation and maintenance of scooters.
Gig workers were mainly offered two types of tasks: collecting and charging scooters overnight, and repairing scooters.
However, scooter ridesharing companies are now ditching gig workers for real employees due to the controversial behaviors performed by gig workers. 
For example, some scooter handlebars and wheels were deliberately damaged to create a chance to be paid to patch them up.
Some scooters were hidden and let the battery die to reap a large payout.
Also, we observed many negative words (e.g., abuse, sturdier, and ditch) under the topic of Gig Worker in Table~\ref{tab:topic_model}.
We only identified four photos depicting the gig jobs in our randomly selected 1770 (10\% of all images) image corpus, which might indicate the decreasing popularity of the gig jobs.

\subsection{Scooter Companies}
\begin{figure}[ht]
  \includegraphics[width=\linewidth]{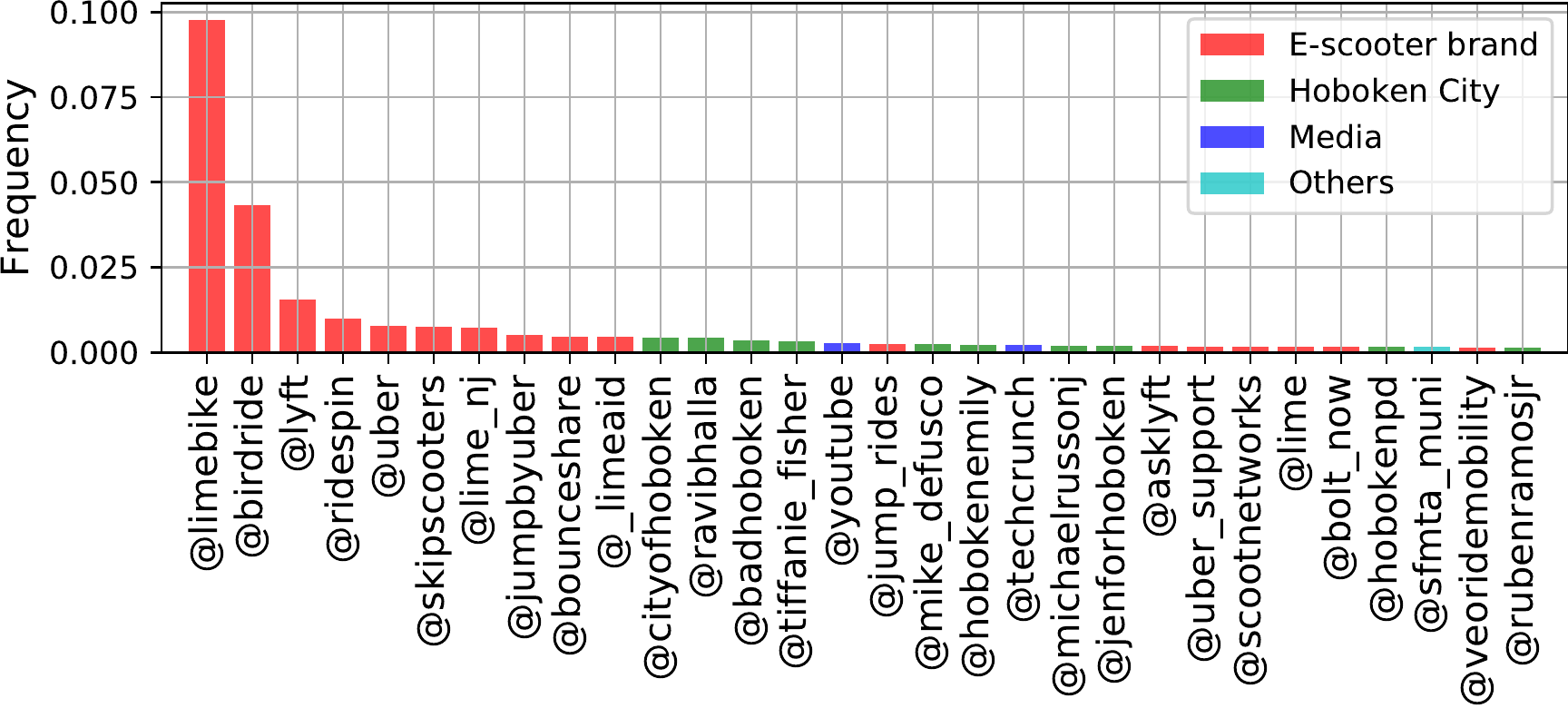}
  \caption{The 20 most frequently mentioned Twitter accounts
  \label{fig:mentioned}}
\end{figure}

Along with the popularity of micro-mobility services, e-scooter operators compete for customers.
Since Twitter serves as a new channel for customer support, we first studied the distribution of @mentioned accounts in our collected dataset.
As shown in Figure~\ref{fig:mentioned}, all of the top 10 most frequent mentions are e-scooter ridesharing brands.
Note that \textit{Jump} scooters are operated by \textit{Uber}.
The Twitter accounts \textit{@limebike}, \textit{@birdride}, and \textit{@lyft} account for more than 15.6\% of all mentions, corresponding to the brands of \textit{Lime}, \textit{Bird}, and \textit{Lyft}.

\begin{figure}[ht]
  \includegraphics[width=\linewidth]{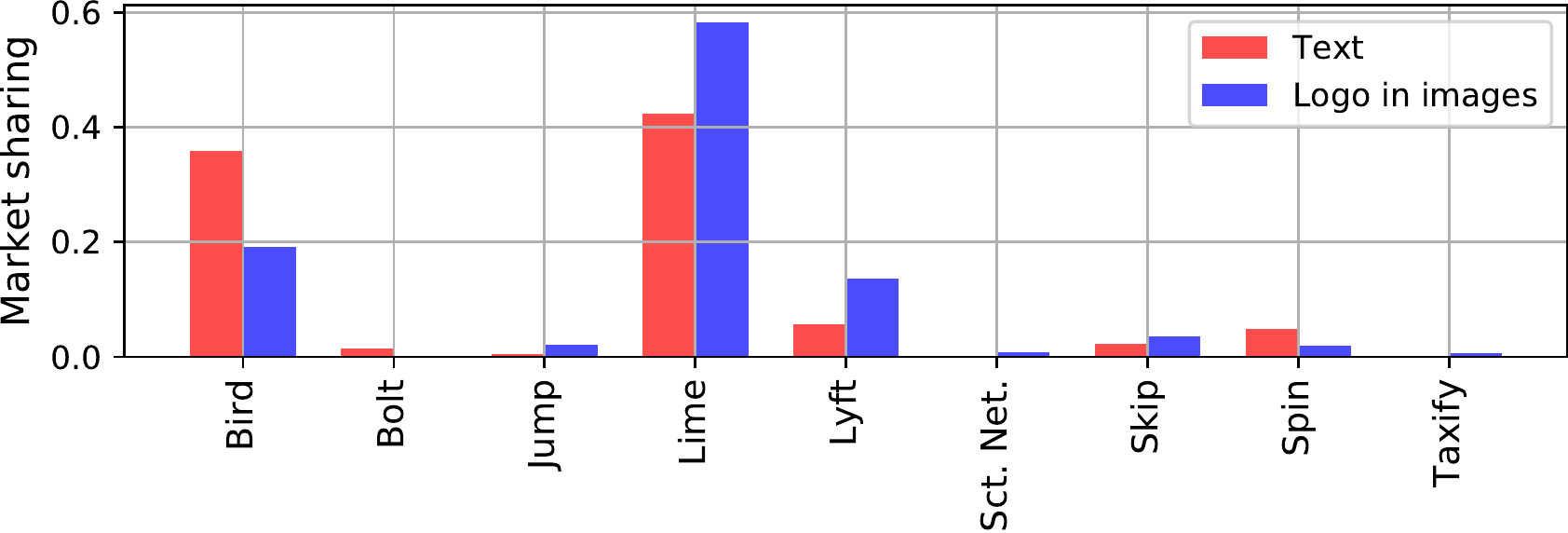}
  \caption{Market shares based on tweet text and logo detection
  \label{fig:market_sharing}}
\end{figure}

We then analyzed the market share of e-scooter ridesharing competitors using both tweet text and posted images.
As a list of scooter brands was applied to reduce false positive during data cleaning, we focused on the same companies when exploring market shares in the e-scooter sharing business.
For the tweet text based analysis, frequencies of company names appearing in tweets were aggregated to estimate their proportions.
The results are demonstrated in Figure~\ref{fig:market_sharing} (see the red bars).
Note that those brands with a market share below 0.5\%, such as \textit{VeoRide}, are not illustrated in Figure~\ref{fig:market_sharing}.

We also recognized and extracted product logos from images to evaluate market shares. 
When processing images, we observed that a large number of duplicated images existed even though retweets had been removed during data preprocessing.
Therefore, we utilized dHash~\cite{dhash}, an image fingerprint hash function, to find duplicated images, and obtained 10,132 unique images after removing duplicates.
Then, we called Google Cloud Vision Logo Detection APIs to detect the involved logos.
The e-scooter sharing logos are summarized in Figure~\ref{fig:market_sharing} (see the blue bars).
Although market shares based on tweet text and extracted logos are different, the top three brands, i.e., \textit{Lime}, \textit{Bird}, and \textit{Lyft}, are in agreement with the top @mentioned Twitter accounts in Figure~\ref{fig:mentioned}.

\section{Operations}
In this section, we investigated four common topics in e-scooter daily operations: products, transactions, parking places, and injuries.
For each topic, we reported the insights and patterns founded through natural language processing and computer vision techniques.

\subsection{Products}
Inspired by the consisting words of the product topic in Table~\ref{tab:topic_model}, we summarized three types of people's concerns about e-scooter product designs.
First, people cared about the usability (perhaps for riders) and portability (perhaps for chargers) of e-scooter, such as self-balance ability, sizes, and foldability.
Second, accessories like Bluetooth, LED, speakers, and lights were extensively discussed by customers.
Except for Bluetooth, all above accessories help improve the riding safety.
Third, the designed speed of e-scooters was of interest to users, as they mentioned the words of \textit{max}, \textit{roller}, \textit{motor}, \textit{mini}, and \textit{spinner} under the product topic.

\subsection{Transactions}
Transactions of e-scooter ridesharing services must be conducted on smartphone apps.
Riders first download scooter apps, sign up, and input payment information (e.g., credit card numbers).
Then they scan a code attached on e-scooters to unlock the scooter for a trip.
After finishing the trip, users are charged and the scooter is locked automatically.
We observed some twitter users shared the app screenshots of the transaction summary page where the payment and trip duration were shown. 
We utilized the Google Cloud Vision Optical Character Recognition (OCR) APIs to extract the trip information from the posted screenshots.
Specifically, 589 unique images containing the dollar sign ``$\$$'' were identified automatically.
Among them, 133 images were recognized as e-scooter app screenshots manually.
Then we designed regular expressions to extract the payment amount and corresponding trip duration from the screenshot OCR results. 
Finally, 78 pairs of payments and riding duration records were found.

The median payment and median duration were \$3.8 and 15.0 minutes, which were close to the average \$3.5 and 16.4 minutes per trip reported by the National Association of City Transportation Officials (NACTO)~\cite{NACTO-report}.
The average payment and duration in our study were \$8.9 and 44.3 minutes with standard deviations of 14.6 and 95.7 respectively.
We think it was caused by the failure or forgetting to lock scooters after finishing the trip.
For example, we observed one 2.4-mile trip lasted 461 minutes and cost \$70.15, and another 0.3-mile trip lasted 426 minutes and cost \$64.90.

\subsection{Injuries}
Injuries associated with shared e-scooters have drawn great attention in recent years.
We found 153 self-reported injury related photos in our collected images, falling into three categories, namely head (22.88\%), trunk \& hands (27.45\%), and legs \& foot (49.67\%), as illustarted in Table~\ref{tab:injury}.
Legs \& foot related injuries were almost twice as likely to occur as that for head or trunk \& hands.
We further divided each type of injuries into five subcategories.
Knee (24.84\%) and hand (11.76\%) were the two most vulnerable body part when riding e-scooters.
Heel injury (0.65\%) and finger injury (2.61\%) were among the least common wound types.
As to head part, chin, eye, mouth and nose were at the same level of vulnerability with the injury ratio range between 5.23\% to 7.19\%.
We noticed the more than half of chin related injuries were very severe.

Lessons on wearing appropriate protective equipment we learned from the above findings can be summarized as follows.
First, knee protective gears are required because of their highest injury frequency in all body parts.
Second, fingerless gloves can be a good choice for riders to avoid hand bruises, the second most common injuries, and enable touching smartphone screens at the same time.
Third, a helmet with chin protection is a must because over half of the reported chin wounds were very serious.
We believe the above three suggestions could be utilized to improve the safety of riders. 

\begin{table*}[]
\setlength{\tabcolsep}{5pt}
\centering 
\caption{Self-reported injury categories} 
\label{tab:injury} 
\scalebox{1.0}{ 
\begin{tabular}{@{}ccccc|ccccc|ccccc@{}}
\toprule
\multicolumn{5}{c|}{Head (22.88\%)}           & \multicolumn{5}{c}{Trunk \& Hands (27.45\%)} & \multicolumn{5}{|c}{Leg \& Feet (\textbf{49.67\%})}             \\ \midrule
Chin   & Eye    & Mouth  & Nose   & Others   & Arm     & Elbow  & Finger & Hand    & Others  & Ankle   & Heel   & Knee    & Thigh  & Others   \\ 
5.23\% & 7.19\% & 5.88\% & 5.88\% & \textbf{11.76\%} & 8.50\%  & 5.88\% & 2.61\% & \textbf{11.76\%} & 0.65\% & 11.11\% & 0.65\% & \textbf{24.84\%} & 3.92\% & 10.46\% \\ \bottomrule
\end{tabular}}
\end{table*}

\subsection{Parking Behaviors}
\begin{figure}[ht]
\subfigure[Under water
  \label{fig:under_water}]{\includegraphics[width=0.3\linewidth]{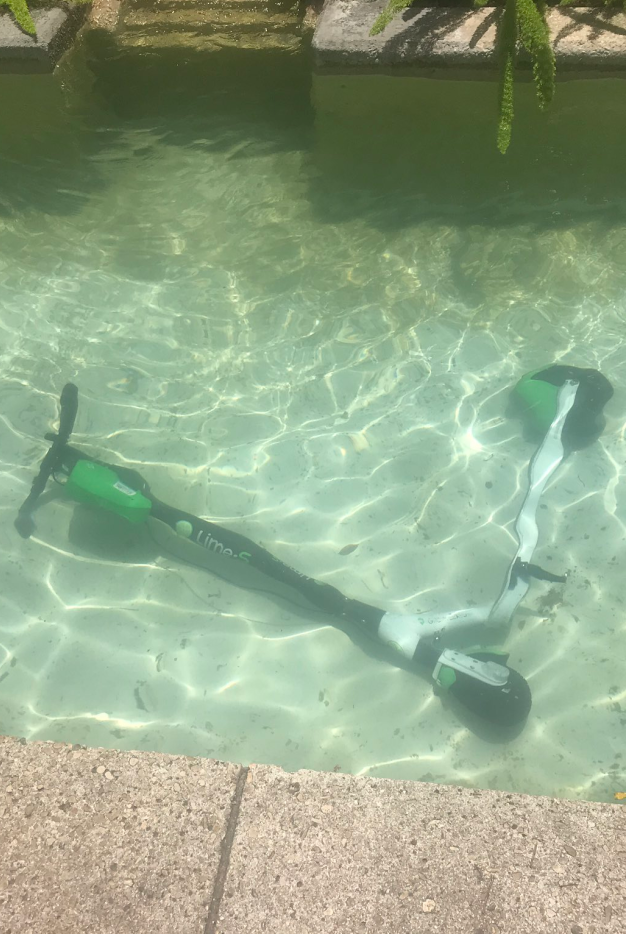}}
\subfigure[On fire
    \label{fig:on_fire}]{\includegraphics[width=0.3\linewidth]{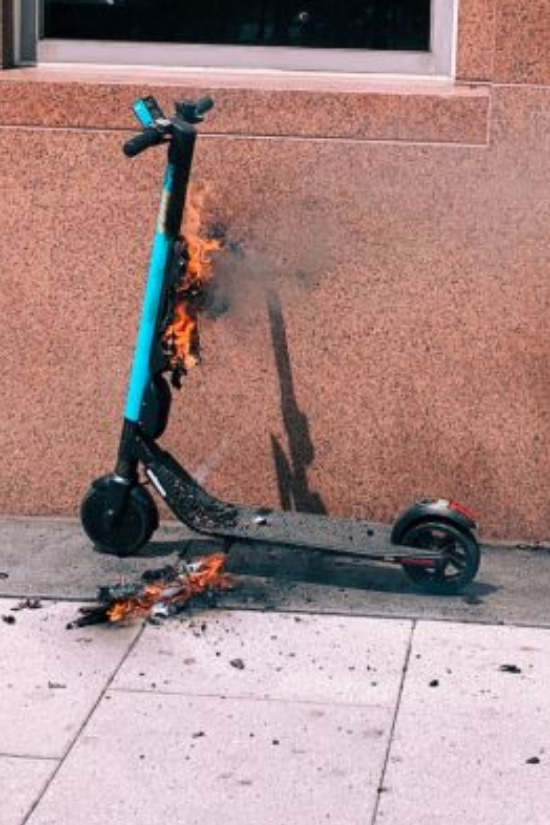}}
\subfigure[Up in trees
  \label{fig:up_in_trees}]{\includegraphics[width=0.3\linewidth]{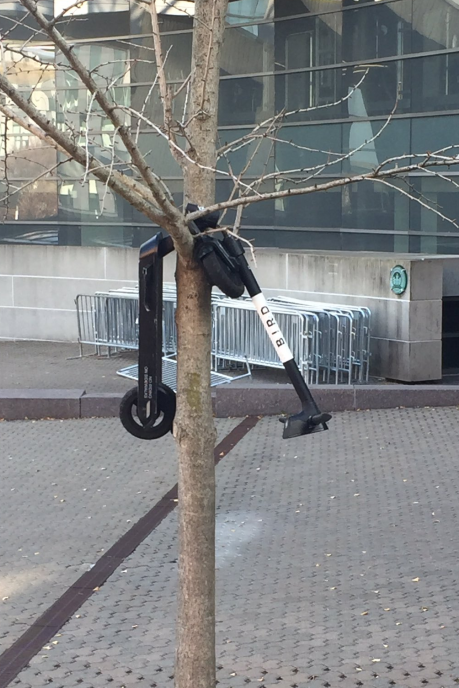}}  
  \caption{Vandalism examples}
  \label{fig:vandalism_examples}
\end{figure}

We randomly selected 10\% collected images and analyzed scooter parking patterns.
Specifically, we identified 230 unduplicated scooter parking images from 1770 images.
We found 37.39\% e-scooters were docked at right places properly such as e-scooter exclusive parking spots, and the rest 62.61\% were in wrong places.
Among those e-scooters parked improperly,
the vast majority of scooters -- 34.78\% of the overall total -- were parked in the middle of sidewalks; 4.78\% were placed indoors; 5.65\% were vandalized; and 17.39\% were parked in other wrong areas.
Unsurprisingly, blocking sidewalks was the most common improper e-scooter parking behavior.
For e-scooters parked indoors, they were found in car parking garages, elevators, and even restrooms.
In addition, we observed more cases of vandalism than indoor parking.
Figure~\ref{fig:vandalism_examples} shows three vandalism examples of under-water, on-fire, up-in-trees scooters.

\begin{figure*}[ht]
  \centering
  \subfigure[Positive word cloud
  \label{fig:wordcloud_positive}]{\includegraphics[width=0.32\linewidth]{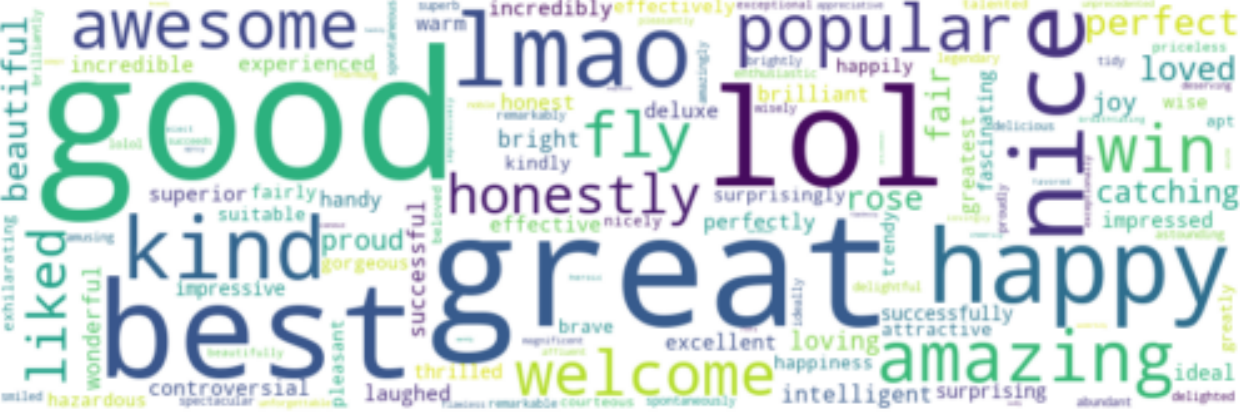}}
  \subfigure[Negative word cloud
  \label{fig:wordcloud_negative}]{\includegraphics[width=0.32\linewidth]{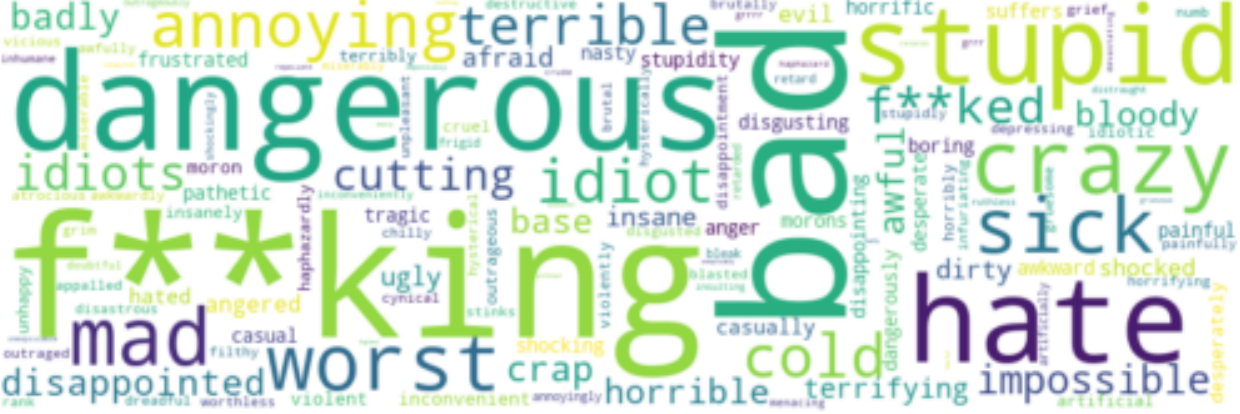}}
  \subfigure[Positive and negative word cloud
  \label{fig:wordcloud_polarized}]{\includegraphics[width=0.32\linewidth]{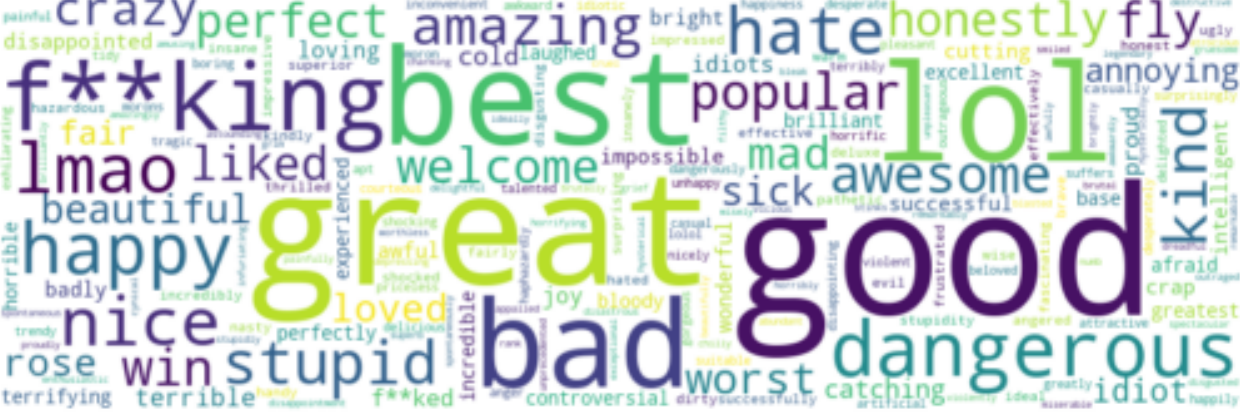}}
  \caption{Polarized word cloud. Positive words with a polarity larger than 0.5 and negative words with a polarity less than -0.5 in TextBlob.
  \label{fig:word_cloud}}
\end{figure*}

\section{Emotion Analysis}
We analyze the emotions of Twitter users who discussed shared e-scooters from three angles, i.e., polarized words, facial emojis, and emoticons. 

\subsection{Polarized Words}
Several lexicons were used to analyze polarized words. 
First, we followed the Twitter-specific lexicons proposed in~\cite{naveed2011bad}.
We found that the most common positive terms included ``like'' (n=17,407), ``great'' (n=4176), ``excellent'' (n=161) and ``rock on'' (n=10), and the most common negative terms included ``f**k'' (n=4426), ``fail'' (n=1137) and ``suck'' (n=588), ``eww'' (n=98).
The total number of positive words is 3.48 times as many as negative ones.

Then, to take more polarized words into account, we created a large polarized word corpus using the TextBlob~\cite{loria2014textblob}.
For each given term, TextBlob returns a polarity score ranging in $[-1.0, 1.0]$, where $-1.0$ is the most negative, $1.0$ is the most positive, and $0.0$ is neutral.
We used polarity thresholds $0.5$ and $-0.5$ to identify terms with confident polarity.
In other words, a term with a TextBlob polarity score above $0.5$ is considered positive, and negative if its TextBlob polarity score is below $-0.5$. 
Figure~\ref{fig:word_cloud} shows the word clouds of the polarized words we detected.
We can see those words ``great,'' ``good,'' ``best,'' and initialisms like ``LOL'' (laughing out loud) and ``LMAO'' (laughing my ass off) are frequently used to express their positive experiences.
In Figure~\ref{fig:word_cloud}(b), we noticed that ``f**king,'' ``bad,'' ``dangerous,'' are highlighted, indicating users' concerns about shared e-scooter.
Figure~\ref{fig:word_cloud}(c) demonstrates that positive emotions dominate the whole polarized words. 

\subsection{Facial Emojis}
The facial emojis are officially classified into positive, neutral, and negative sentimental groups by the Unicode Consortium.
We summarized the top 15 most frequent facial emojis in each group (62.6\% positive, 16.2\% neutral, and 21.2\% negative) in Table~\ref{tab:emoji}.
The most widely used emoji in this study is the face with tears of joy~\includegraphics[width=0.023\linewidth]{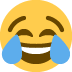}, which is also reported as the most popular emoji globally~\cite{meaning-of-face-with-tears-of-joy}.
The money-mouth face emoji~\includegraphics[width=0.023\linewidth]{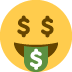} might demonstrate that e-scooter sharing services are affordable and cost-effective.
For neutral sentiments, only eleven types of facial emojis were mentioned.
In negative emojis, skull emoji~\includegraphics[width=0.023\linewidth]{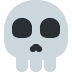}, exploding head emoji~\includegraphics[width=0.023\linewidth]{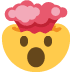},
and the face with head-bandage emoji~\includegraphics[width=0.023\linewidth]{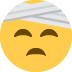} indicated the possible dangers and injuries when riding e-scooters.
Also, people used the face with symbols on mouth emoji~\includegraphics[width=0.023\linewidth]{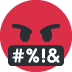} and pouting face emoji~\includegraphics[width=0.023\linewidth]{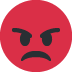} to express their anger on shared e-scooters.

\subsection{Emoticon}
Although emojis are gradually replacing emoticons on social media, emoticons are still ubiquitous because of their simplicity and platform independence.
We summarized the usage of all positive and negative emoticons in the list of sideways Latin-only emoticons~\cite{emoticon-list}, as shown in Table~\ref{all-emoticons}.
The most popular positive emoticons in this study are ``:)'' (n=490), ``:3'' (n=467), and ``;)'' (n=137), while the most popular negative ones are mainly expressed by ``:/'' (n=325), ``:('' (n=321), and ``:$\backslash$'' (n=325). 
Similar to both polarized words and facial emojis, the positive emotions override the negative ones. 
We concluded that most people embraced this novel mode of micromobility but with reasonable concerns.

\begin{table*}[h] 
 \setlength{\tabcolsep}{1pt}
 \centering 
 \caption{Top emojis by sentiment categories (numbers represent frequency)} 
 \label{tab:emoji} 
 \begin{tabular}{l|ccccccccccccccccc} 
  \toprule
  
Positive & \includegraphics[width=0.05\linewidth]{figures/emoji_image/1f602.png} & \includegraphics[width=0.05\linewidth]{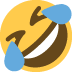} & \includegraphics[width=0.05\linewidth]{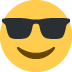} & \includegraphics[width=0.05\linewidth]{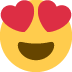} & \includegraphics[width=0.05\linewidth]{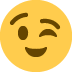} & \includegraphics[width=0.05\linewidth]{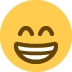} & \includegraphics[width=0.05\linewidth]{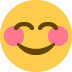} & \includegraphics[width=0.05\linewidth]{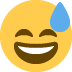} & \includegraphics[width=0.05\linewidth]{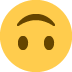} & \includegraphics[width=0.05\linewidth]{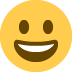} & \includegraphics[width=0.05\linewidth]{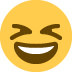} & \includegraphics[width=0.05\linewidth]{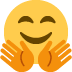} & \includegraphics[width=0.05\linewidth]{figures/emoji_image/1f911.png} & \includegraphics[width=0.05\linewidth]{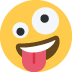} & \includegraphics[width=0.05\linewidth]{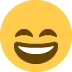} &
\includegraphics[width=0.05\linewidth]{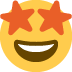} & \\ 

& 942 & 212 & 150 & 122 & 115 & 93 & 78 & 68 & 64 & 64 & 56 & 44 & 41 & 40 & 37 & 37  \\ 

 & \includegraphics[width=0.05\linewidth]{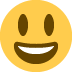} & \includegraphics[width=0.05\linewidth]{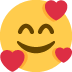} & \includegraphics[width=0.05\linewidth]{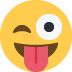} & \includegraphics[width=0.05\linewidth]{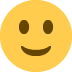} & \includegraphics[width=0.05\linewidth]{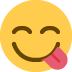} & \includegraphics[width=0.05\linewidth]{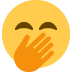} & \includegraphics[width=0.05\linewidth]{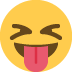} & \includegraphics[width=0.05\linewidth]{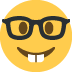} & \includegraphics[width=0.05\linewidth]{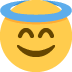} & \includegraphics[width=0.05\linewidth]{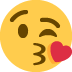} & \includegraphics[width=0.05\linewidth]{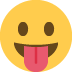} & \includegraphics[width=0.05\linewidth]{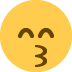} & \includegraphics[width=0.05\linewidth]{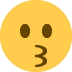} & \includegraphics[width=0.05\linewidth]{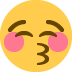} & \includegraphics[width=0.05\linewidth]{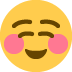} &   \\ 

& 32 & 30 & 21 & 19 & 18 & 15 & 15 & 12 & 11 & 11 & 7 & 2 & 1 & 1 & 1 & \\  \midrule 
Neutral & \includegraphics[width=0.05\linewidth]{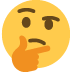} & \includegraphics[width=0.05\linewidth]{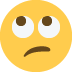} & \includegraphics[width=0.05\linewidth]{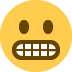} & \includegraphics[width=0.05\linewidth]{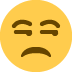} & \includegraphics[width=0.05\linewidth]{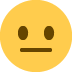} & \includegraphics[width=0.05\linewidth]{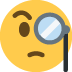} & \includegraphics[width=0.05\linewidth]{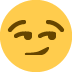} & \includegraphics[width=0.05\linewidth]{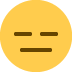} & \includegraphics[width=0.05\linewidth]{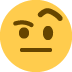} & \includegraphics[width=0.05\linewidth]{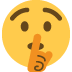} & \includegraphics[width=0.05\linewidth]{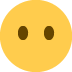} &  &  &  &  &  \\ 
 & 224 & 110 & 41 & 35 & 33 & 31 & 25 & 22 & 19 & 5 & 4 &  &  &  &  & \\ \midrule 

Negative & \includegraphics[width=0.05\linewidth]{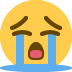} & \includegraphics[width=0.05\linewidth]{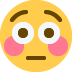} & \includegraphics[width=0.05\linewidth]{figures/emoji_image/1f480.png} & \includegraphics[width=0.05\linewidth]{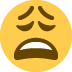} & \includegraphics[width=0.05\linewidth]{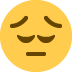} & \includegraphics[width=0.05\linewidth]{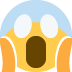} & \includegraphics[width=0.05\linewidth]{figures/emoji_image/1f92c.png} & \includegraphics[width=0.05\linewidth]{figures/emoji_image/1f621.png} & \includegraphics[width=0.05\linewidth]{figures/emoji_image/1f92f.png} & \includegraphics[width=0.05\linewidth]{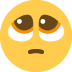} & \includegraphics[width=0.05\linewidth]{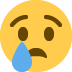} & \includegraphics[width=0.05\linewidth]{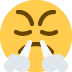} & \includegraphics[width=0.05\linewidth]{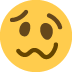} & \includegraphics[width=0.05\linewidth]{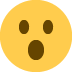} & \includegraphics[width=0.05\linewidth]{figures/emoji_image/1f915.png} & \includegraphics[width=0.05\linewidth]{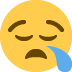} & \\
 & 183 & 70 & 69 & 58 & 43 & 38 & 36 & 35 & 32 & 31 & 30 & 27 & 24 & 22 & 21 & 15 \\ 
& \includegraphics[width=0.05\linewidth]{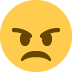} & \includegraphics[width=0.05\linewidth]{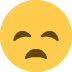} & \includegraphics[width=0.05\linewidth]{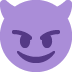} & \includegraphics[width=0.05\linewidth]{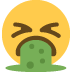} & \includegraphics[width=0.05\linewidth]{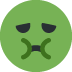} & \includegraphics[width=0.05\linewidth]{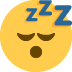} & \includegraphics[width=0.05\linewidth]{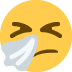} & \includegraphics[width=0.05\linewidth]{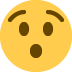} & \includegraphics[width=0.05\linewidth]{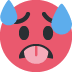} & \includegraphics[width=0.05\linewidth]{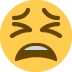} & \includegraphics[width=0.05\linewidth]{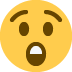} & \includegraphics[width=0.05\linewidth]{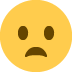} & \includegraphics[width=0.05\linewidth]{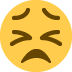} & \includegraphics[width=0.05\linewidth]{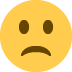} & \includegraphics[width=0.05\linewidth]{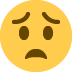} & \includegraphics[width=0.05\linewidth]{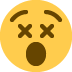} \\ 
 & 13 & 12 & 12 & 12 & 11 & 11 & 11 & 10 & 9 & 9 & 9 & 8 & 8 & 8 & 7 & 7 & \\ \bottomrule 
 \end{tabular}
 \end{table*}

\begin{table*}[ht] 
\vspace{0.5cm}
\setlength{\tabcolsep}{6pt}
\centering 
\caption{Top emoticons by sentiment categories (numbers represent frequency)} 
\label{all-emoticons} 
    \begin{tabular}{l|ccccccccccccccc} 
        \toprule
        Positive & :) 490 & :3 467 & ;) 137 & :-) 87 & 8) 79 & :p 35 & ;-) 33 & :b 16 & *) 9 & :* 8 & =) 7 & :$\hat{\mkern6mu}$) 6 & x-p 6 & :-)) 4 & :] 3 \\ \midrule
        Negative & :/ 325 & :( 321 & :$\backslash$ 245 & :-( 22 & :o 18 & :-/ 14 & :\$ 10 & :c 9 & 8-0 6 & :[ 3 & :\{ 2 & =$\backslash$ 2 & :-c 0 & :-< 0 & :< 0 \\ \bottomrule 
    \end{tabular}
\end{table*}

\section{Conclusion}
In this paper, we leveraged massive volumes of heterogeneous Twitter data, including text, @mentions, GPS data, general photos, screenshots of e-scooter apps, emojis, and emoticons, to study e-scooter ridesharing services on a large scale.
More than five million English tweets mentioning the word ``scooter'' or the scooter emoji were collected continuously in 18 months.
We first performed a comprehensive data preprocessing to remove noises and reduce false-positive effects.
Then, we profiled the geospatial and temporal tweet distributions with multiple granularities.
The underlying popular topics were summarized using both \#hashtags and the LDA topic model.
For each of the extracted topics, we reported the profound insights and patterns, such as the popularity in different cities, the gender gap of riders, e-scooter market shares, transaction information, injury types, parking behaviors, and emotions from the public.
We believe the crowdsourced findings provide a deep understanding of the emerging shared e-scooter services in smart cities. 

\bibliographystyle{IEEEtran}
\bibliography{bibliography}

\end{document}